# Lambda- and Schottky-anomalies in solid-state phase transitions

Yuri Mnyukh
*76 Peggy Lane, Farmington, CT, USA, e-mail: yuri@mnyukh.com*
(Dated: April 24, 2011)

The origin of lambda and Schottky anomalies in solid-state phase transitions are analyzed and illustrated. They are shown to be the latent heat of nucleation-and-growth phase transitions.

## 1. Introduction

The sharp peaks of heat capacity reminiscent to letter λ, recorded at the temperatures of solid-state phase transitions, challenged theorists to explain their origin. The first λ-peak was observed by Simon in $NH_4Cl$ phase transition [1]. Later on, many other cases were reported. Thus, more than 30 experimental λ-peaks presented as "Specific heat $C_P$ of [substance] *vs.* temperature *T*" were shown in the book by Parsonage and Staveley [2]. The existed theories turned out to be unable to account for the phenomenon. P.W. Anderson wrote [3]: "Landau, just before his death, nominated [lambda-anomalies] as the most important as yet unsolved problem in theoretical physics, and many of us agreed with him...Experimental observations of singular behavior at critical points...multiplied as years went on...For instance, it have been observed that magnetization of ferromagnets and antiferromagnets appeared to vanish roughly as $(T_C-T)^{1/3}$ near the Curie point, and that the λ-point had a roughly logarithmic specific heat $(T-T_C)^0$ nominally". Feynman stated [4] that "One of the challenges of theoretical physics today is to find an exact theoretical description of the character of the specific heat near the Curie transition - an intriguing problem which has not yet been solved."

There were three main reasons for that theoretical impasse. (1) The search was limited by the framework of second-order phase transitions, while the λ-peaks were actually observed in first-order phase transitions. (2) The first-order phase transitions exhibiting the λ-peaks were treated as second order; the latent heat was lost there. As shown below, the real mechanism of first order phase transitions must be known in order to explain those peaks. (3) An important limitation of the calorimetry utilized in the measurements was unnoticed.

## 2. Nucleation-and-growth phase transitions

The molecular mechanism of first-order solid-state phase transitions has been revealed in detail in the studies [5-18] summarized in [19]. It is a *crystal growth* involving nucleation and propagation of interfaces. Importantly, it covers ferromagnetic and ferroelectric phase transitions as well [19-24].

The *nucleation* is a key. It is not the classical fluctuation-based process described in textbooks. In a given crystal it is pre-determined. The nuclei are located in specific crystal defects - microcavities of a certain optimum size. These defects already contain information on the temperatures $T_n$ of their activation. Any nucleation act is followed by the molecule-by-molecule crystal rearrangement at the interface. The nucleation lags $\Delta T_n = T_n - T_o$ (at $T_o$ the free energies of the polymorphs are equal) are inevitable and reproducible for a given defect, but are not the same in different defects. Considering that almost all real systems (polycrystals, imperfect or polydomain crystals, etc.) feature multiple nucleation, the phase transition is spread over a temperature *range of transition;* this range is a subject to *hysteresis.* At any fixed intermediate temperature the system of the two coexisting phases is in a quasi-equilibrium state. A slow temperature change alters the mass fractions of the low- and high-temperature phases $m_L$ and $m_H$ in the two-phase system $m_L+m_H = 1$. In a sense, first-order phase transitions are also continuous, but this time it is *quantity* rather than *quality* that changes continuously with temperature. The "jumps" of their physical properties, known to be their main feature, appear as such on the recordings only when the range of transition is narrow and/or passed quickly. No macroscopic "jumps" actually occur during the phase transition. They are simply the differences between physical properties of the initial and resultant phases.

## 3. Analyzing old literature data

The canonical case of "specific heat λ-anomaly" in $NH_4Cl$ around -30.6 °C will be re-examined. This case is of a special significance. It was the first where a λ-peak in specific heat measurements through a solid-state phase transition was reported [1]. It was the only example used by Landau in his original articles on the theory of continuous second-order phase transitions [25]. This phase transition was a subject of numerous studies by different experimental techniques and considered most thoroughly investigated. In every calorimetric work (*e.g.*, [26-34] ) a sharp λ-peak in this phase transition was recorded; neither author expressed doubts in a specific heat nature of the peak. The transition has been designated as a *cooperative order-disorder phase transition of the lambda type* and used to exemplify such a type of phase transitions. However, no one maintained that the λ-anomaly was well understood.

It should be noted that many of the above-mentioned calorimetric studies were undertaken well after the experimental work by Dinichert [35] was published in 1942. It revealed (Fig. 1a) that the transition in *NH₄Cl* was spread over a temperature range where only mass fractions $m_L$ and $m_H$ of the two distinct *L* (low-temperature) and *H* (high-temperature) coexisting phases were changing, producing "sigmoid"-shaped curves. The direct and reverse runs formed a hysteresis loop. The fact that the phase transition was of the first-order was incontrovertible.

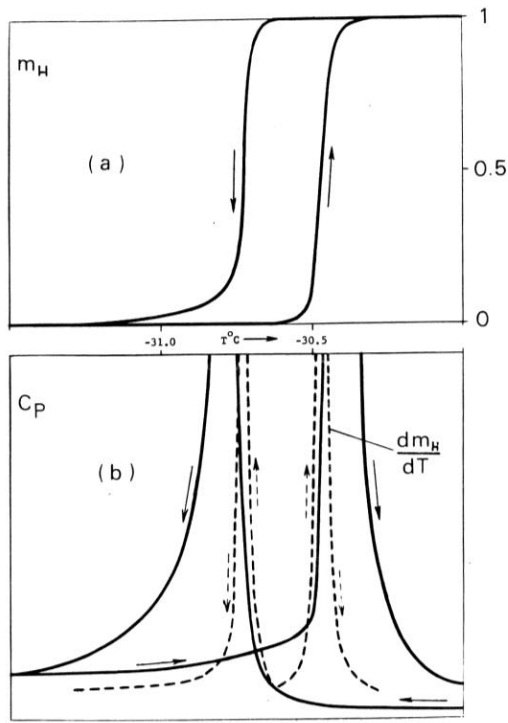

Fig. 1. Phase transition in *NH₄Cl*.
(a) The hysteresis loop by Dinichert represents mass fraction of high-temperature phase, $m_H$, in the two-phase, *L+H*, range of transition; $m_L + m_H = 1$.
(b, solid lines) The λ-peaks from calorimetric measurements by Extermann and Weigle.
The plots are positioned under one another in the same temperature scale to make it evident that the shape of the peaks is proportional to *fist derivative* (dotted curves) of the $m_H(T)$.

In Fig. 1 the Dinichert's (x-ray) data are compared with the calorimetric measurements by Extermann and Weigle [28]. The latter exhibited "anomalies of heat capacity", as the authors called the λ-peaks, and the hysteresis of these peaks. (At this point, however, it had to become evident that the λ-peaks cannot be of a heat capacity nature, considering that heat capacity is a *unique function* of temperature). The graphs 'a' and 'b' are positioned under one another in the same temperature scale to reveal that the shape and location of the peaks are very close to *fist derivative* (dashed curves) of the $m_H(T)$. It remains only to note that *latent heat* of the phase transition must be proportional to $dm_H/dT$. Thus, the latent heat of the first-order phase transition, lost in the numerous calorimetric studies, is found. Fig. 2 sums up the foregoing dissection.

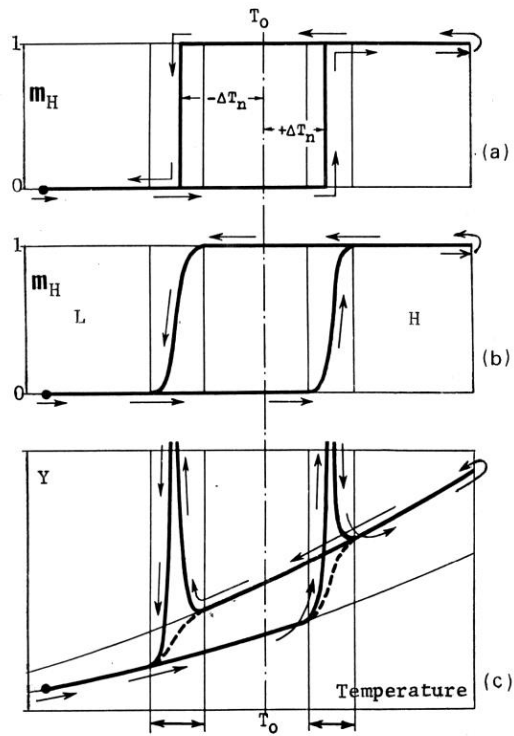

Fig. 2. Formation of latent heat λ-peaks in a phase transition. "Sigmoid" curves, hysteresis loop, and λ-peaks upon a phase transition cycle $L \rightarrow H \rightarrow L$. $m_H$ is mass fraction of *H* phase. $T_o$ is temperature such that the free energies of the phases are equal. Two-phase ranges of transitions are marked by bilateral arrows.
(a) Hysteresis of a single small particle in a powder or polycrystalline specimen.
(b) Spread of each transition over a temperature range for the specimen as a whole owing to different nucleation temperatures $T_n$ in the constituent particles. Two continuous "sigmoid" curves, one in the forward and the other in the reverse direction, form a hysteresis loop.
(c) Total calorimetric output *Y* comprising the heat capacities $C_P(T)$ of the two phases subject to their mass fractions $C_L m_L + C_H m_H$ (dotted "sigmoid" curves) and the latent heat represented by the area under the λ-peaks superimposing the "sigmoid" curves.

### 4. Limitations of adiabatic calorimetry

A legitimate question can be raised: why did not publication of the Dinichert's work change the λ-peaks interpretation from "heat capacity" to "latent heat"? The answer is: knowledge of the actual phase transition mechanism outlined in section 2 was required. But there was also a secondary reason hidden in the calorimetric technique itself.

The goal of numerous calorimetric studies of λ-peaks in $NH_4Cl$ and other substances was to delineate shape of these peaks with the greatest possible precision. An adiabatic calorimetry, it seemed, suited best to achieve it. The adiabatic calorimeters, however, are only "one way" instruments in the sense the measurements can be carried out only as a function of increasing temperature. In the case under consideration, however, it was vital to perform both temperature-ascending and descending runs - otherwise existence of hysteresis would not be detected. And it was not detected. For example, in [33] the transition in $NH_4Cl$ was interpreted as occurring at the fixed temperature point $T_\lambda = 245.502 \pm 0.004\ ^\circ K$ defined as a position of λ-peak. The high precision of measurements was useless: that $T_\lambda$ exceeded $T_o$ by $3^\circ$.

The results by Extermann and Weigle were not typical. The kind of calorimetry they utilized permitted both ascending and descending runs. That was a significant advantage over the adiabatic calorimetry used by others in the subsequent years. But there was also a shortcoming in their technique resulted in the unnoticed error in the presentation of the λ-peaks in Fig 1b: the peak in the descending run had to be *negative* (looking downward). The error was purposely not corrected in Fig. 2 to consider it separately.

### 5. Examination by differential scanning calorimetry (DSC).

DSC is free of the above shortcomings. Carrying out temperature descending runs with DSC is as easy as ascending runs. Most importantly, it displays endothermic and exothermic peaks with *opposite* signs in the chart recordings, which results from the manner the signal is measured [36]. If the λ-peak in $NH_4Cl$ is a *latent heat* of phase transition, as was concluded above, the peak in a descending run must be exothermic and look downward. Our strip-chart recordings made with a Perkin-Elmer DSC-1B instrument immediately revealed that the peak acquires opposite sign in the reverse run (Fig. 3). Its hysteresis was also displayed.

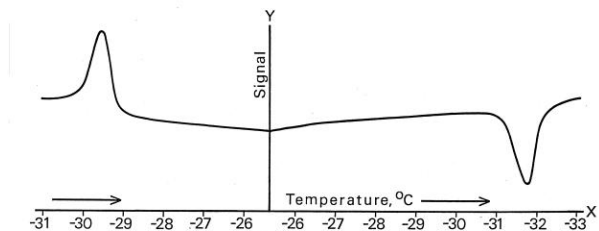

Fig. 3. The actual DSC recording of $NH_4Cl$ phase transition cycle, displaying temperature-ascending and descending peaks as being endothermic and exothermic accordingly, thus delivering final proof of a latent heat nature of the λ-peak.

### 6. Schottky-anomalies

Whether an "anomaly" is *lambda* or *Schottky* is determined by (1) the latent heat value and (2) the length of the range of transition. If the latent heat is emanated over a short temperature range, it may look like a λ-peak, but if the same latent heat is spread over a long temperature range, it will look like a hump and called a Schottky-anomaly. The range of transition is narrower in layered structures. It differs widely for the same substance depending whether the sample is a single crystal, polydomain crystal, polycrystal, or fine powder.

### 7. λ-Peaks of other physical properties

The belief in the heat capacity λ-anomalies in solid-state phase transitions was seemingly supported by finding that some other physical properties also exhibit analogous peaks. It remained unknown or underestimated that those phase transitions were of first order (consequently, nucleation-and-growth). There are several different causes for these peaks to appear. They are considered in detail in Chapter 3 of [19]. Here are three of them.

When passing the two-phase range of transition, the mass fractions of the phases in a sample follow the "sigmoid" curves as shown in Fig 2b. Some physical properties P during that process can be measured only as a combined contribution from the two phases in accordance with their current mass fractions $m_H$ and $m_L$. In those cases the observed $P_{obs}(T)$ will have almost the same shape as $m_H(T)$ (or $m_L(T)$). It remains to note that a *first derivative* of any "sigmoid" curve is a peak. Therefore, any physical property defined as $dP(T)/dT$ will exhibit peak. The typical such property is a volume coefficient of thermal expansion $\alpha$, which by definition is a first temperature derivative of the specific volume $v(T)$, $\alpha = dv(T)/dT$. Fig. 4 illustrates its "λ-anomaly" in $NH_4Cl$. Similarly, the "λ-anomalies" of temperature coefficients of some other properties can be created.

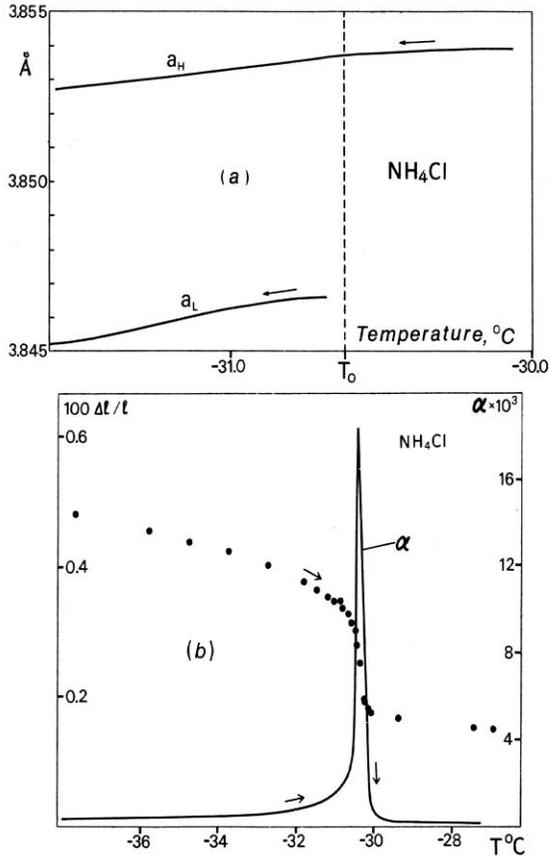

Fig. 4. The conflicting x-ray [35] and dilatometric [37] data through the temperature range of $NH_4Cl$ phase transition.
(a) The lattice parameters $a$ of the cubic unit cells of the $H$ and $L$ phases. The phases independently coexist in a temperature range without gradual change from $a_H$ to $a_L$ (and hence the unit cell volume from $(a_H)^3$ to $(a_L)^3$). The difference was accountable for 0.45% divergence in the crystal densities.
(b) The array of experimental points is the dilatometric relative linear thermal expansion $\Delta\ell/\ell$ in $NH_4Cl$. The solid line is a λ-peak of the volume coefficient $\alpha$ of thermal expansion derived from the dilatometric data. Dilatometry cannot reveal that the continuous change of $\Delta\ell/\ell$ is due to change of the relative content of distinct H and L phases in the specimen.

For example, Rao and Rao [38, p. 295] reproduced an electric resistance vs. temperature plot ("sigmoid" curve) and its derivative (λ-peak) and supplied them with the caption: "Note the second-order transition indicated in the plot of the derivative of resistance".

Another occasional source of "λ-anomaly" illusion is the particular mutual disposition of $P_H(T)$ and $P_L(T)$ shown in Fig. 5.

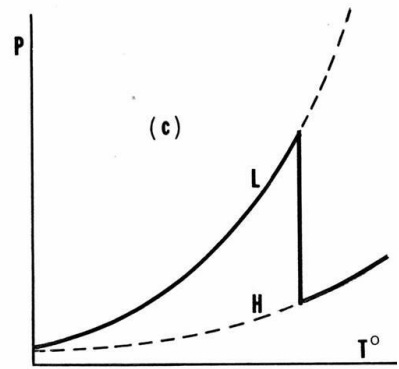

Fig. 5. Illusion of an "λ-anomaly" in case of the shown mutual disposition of $P_H(T)$ and $P_L(T)$. The sketch does not reflect existence of a range of transition which would cause "rounding" of the apparent peak.

Finally, the problem of the narrow peaks of light and neutron scattering at the "critical temperature $T_C$", called *central peak problem* is to be mentioned. The phenomenon was also called *critical opalescence*. The central peaks were observed in $NH_4Cl$ (Fig. 6), quartz, $K_2PO_4$, $SrTiO_3$ and a number of other phase transitions. In time, evidence was mounting, and by 1980 proven, that the peaks were caused by scattering from *static* centers. The issue was presented in detail and referenced in [19] (Sec. 3.6, 3.7 and Appendix entitled *Review on "Light Scattering Near Phase Transitions"*). The static scattering centers were the nuclei and interfaces appearing at the beginning and disappearing at the end of the temperature ranges of nucleation-and-growth phase transitions.

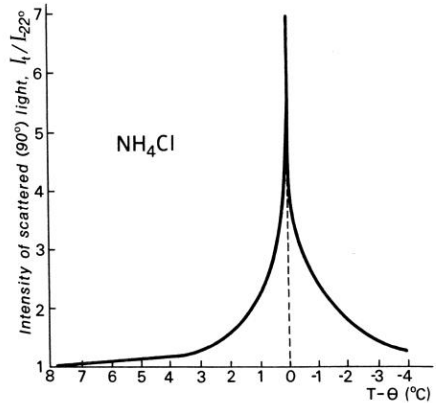

Fig. 6. The apparent "critical opalescence" peak of light scattered by a $NH_4Cl$ single crystal during its phase transition [39]. The temperature scale, presented as distance from the peak position, obscures the fact that the peak itself was not located exactly at $T_o = -30.6\ °C$ owing to hysteresis.

## 7. A note about first-order liquid-liquid phase transitions

Existence of first-order liquid-liquid phase transitions is a well-established fact [40]. To be in compliance with thermodynamics, the only way they can be realized is by nucleation and propagation of interfaces over a range of transition [19, 22]. DSC measurements would inevitably reveal their latent heat λ-peaks.

**References**

[1] F. Simon, *Ann. Phys.* **68,** 241 (1922).
[2] N.G. Parsonage and L.A.K. Staveley, *Disorder in Crystals*, Clarendon Press (1978).
[3] P.W. Anderson, *Science* **218**, no. 4574, 763 (1982).
[4] R.P. Feynman, R.B. Leighton and M. Sands, *The Feynman Lectures on Physics*, v.2, Addison-Wesley (1964).
[5] Y. Mnyukh, *J. Phys. Chem. Solids*, **24** (1963) 631.
[6] A.I. Kitaigorodskii, Y. Mnyukh, Y. Asadov, *Soviet Physics - Doclady* **8** (1963) 127.
[7] A.I. Kitaigorodskii, Y. Mnyukh, Y. Asadov, *J. Phys. Chem. Solids* **26** (1965) 463.
[8] Y. Mnyukh, N.N. Petropavlov, A.I. Kitaigorodskii, *Soviet Physics - Doclady* **11** (1966) 4.
[9] Y. Mnyukh, N.I. Musaev, A.I. Kitaigorodskii, *Soviet Physics - Doclady* **12** (1967) 409.
[10] Y. Mnyukh, N.I. Musaev, *Soviet Physics - Doclady* **13** (1969) 630.
[11] Y. Mnyukh, *Soviet Physics - Doclady* **16** (1972) 977.
[12] Y. Mnyukh, N.N. Petropavlov, *J. Phys. Chem. Solids* **33** (1972) 2079.
[13] Y. Mnyukh, N.A. Panfilova, *J. Phys. Chem. Solids* **34** (1973) 159.
[14] Y. Mnyukh, N.A. Panfilova, *Soviet Physics - Doclady* **20** (1975) 344.
[15] Y. Mnyukh *et al.*, *J. Phys. Chem. Solids* **36** (1975) 127.
[16] Y. Mnyukh, *J. Crystal Growth* **32** (1976) 371.
[17] Y. Mnyukh, *Mol. Cryst. Liq. Cryst.* **52** (1979) 163.
[18] Y. Mnyukh, *Mol. Cryst. Liq. Cryst.* **52** (1979) 201.
[19] Y. Mnyukh, *Fundamentals of Solid-State Phase Transitions, Ferromagnetism and Ferroelectricity*, Authorhouse, 2001 [or 2$^{nd}$ (2010) Edition].
[20] V.J. Vodyanoy, Y. Mnyukh, http://arxiv.org/abs/1012.0967.
[21] Y. Mnyukh, http://arxiv.org/abs /1101.1249.
[22] Y. Mnyukh, http://arxiv.org/abs/1102.1085.
[23] Y. Mnyukh, http://arxiv.org/abs/1103.2194.
[24] Y. Mnyukh, http://arxiv.org/abs/1103.4527.
[25] *Collected Papers of L.D. Landau*, Gordon & Breach (1967).
[26] F. Simon, C.V.Simson, and M. Ruhemann, *Z. Phys. Chem*. **A129,** 339 (1927).
[27] W.T. Ziegler and C.E. Messer, *J. Am. Chem. Soc*. **63,** 2694 (1941).
[28] R. Extermann and J. Weigle, *Helv. Phys. Acta* **15,** 455 (1942).
[29] V. Voronel and S.R. Garber, *Sov. Phys. JETP* **25,** 970 (1967).
[30] W.E. Maher and W.D. McCormick, *Phys. Rev*. **183,** 573 (1969).
[31] D.L. Connelly, J.S. Loomis, and D.E. Mapother, *Phys. Rev.* **B 3**, 924 (1971).
[32] P. Schwartz, *Phys. Rev.* **B 4**, 920 (1971).
[33] H. Chihara and M. Nakamura, *Bul. Chem. Soc. Jap*. **45,** 133 (1972).
[34] J.E. Callanan, R.D. Weir, and L.A.K. Staveley, *Proc. R. Soc. Lond.* **A 372,** 489 (1980); **372,** 497 (1980); **375,** 351 (1981).
[35] P. Dinichert, *Helv. Phys. Acta* **15,** 462 (1942).
[36] J.L. McNaughton and C.T.Mortimer, *Differential Scanning Calorimetry*, Perkin-Elmer, (1975).
[37] A.W. Lawson, *Phys. Rev.* **57**, 417 (1940).
[38] C.N.R. Rao and K.J. Rao, *Phase Transitions in Solids,* McGraw-Hill (1978).
[39] O.A. Shustin, *JETP Letters* **3**, 320 (1966).
[40] Y. Katayama *et al.*, *Nature* **403**, 170 (2000).